\documentclass{osa-article}
\usepackage{bm}
\usepackage{diagbox}
\newcommand{\ubm}[1]{\bm{\hat{#1}}}
\journal{oe}

\begin{document}

\title{Statistic Vectorial Complex Ray Model and its Application to Three-Dimension Scattering of a Non-spherical Particle}

\author{Ruiping YANG,\authormark{1} Bing WEI,\authormark{1,*}  Claude ROZÉ,\authormark{2,*} Saïd IDLAHCEN,\authormark{2}  and  Kuan Fang REN\authormark{1,2}}

\address{\authormark{1}School of physics, Xidian University, Xi'an 710071, Shaanxi Province, People's Republic of China\\
\authormark{2}Univ Rouen Normandie, INSA Rouen Normandie, CNRS, CORIA UMR 6614, F-76000 Rouen, France}

\email{\authormark{*}bwei@xidian.edu.cn, claude.roze@univ-rouen.fr} 




\begin{abstract}
A Statistic Vectorial Complex Ray Model (SVCRM) is proposed for the scattering of a plane wave by a non-spherical dielectric particle in three dimensions. This method counts the complex amplitudes of all rays arriving in a tiny box in the observation direction. It avoids the two-dimensional interpolation necessary in the Vectorial Complex Ray Model (VCRM) for the calculation of the total field. So, it is more flexible and suitable to deal with the particle of complex shape. The algorithm has been carefully designed for the calculation of the phases due to the optical path, the reflection and the focal lines/points as well as the amplitude variation caused by the reflection, the refraction and the divergence of the wave on the particle surface. This model is then applied, as an example, to simulate the three-dimensional scattering patterns of a pendent drop. The scattering mechanism is analyzed in details and the special attention has been paid to the scattering patterns around the rainbow angles where the caustics occur. The simulated results have been compared to the experimental results for a pendent drop of two typical sizes and shapes. It is shown that the simulated and the experimental results are in good agreement. This method opens a promising potential in the development of optical measurement techniques in fluid mechanics.
\end{abstract}

\section{Introduction} \label{sec:Introduction}

Optical metrology is widely employed in various domains of scientific research due to its advantages of being fast and non-intrusive. Numerous measurement techniques \cite{mayinger2013optical,sirohi2018optical,Zhao_2023} have been developed to characterize particle properties, such as the size, the refractive index, the temperature and the velocity. However,  most of these techniques are limited to particles of simple shape because of the lack of a theoretical model for predicting the relationship between the scattered light and the properties of the scatterer, especially for a large non-spherical particle. Various theories and models have been developed to address this problem \cite{mishchenko1999light,MISHCHENKO2009808,2013Electromagneticnon}. They can be categorized into three groups: rigorous theories, numerical methods and approximate models.

The rigorous theories provide exact solutions of the Maxwell equations, such as the Lorenz-Mie theory \cite{hulst1981light, born1959principles} and its generalizations (GLMT) \cite{gouesbet1988light,Barton88,book2017GLMT,xu2007theoretical}. These theories are applicable only to particles of simple shape: sphere, infinite circular cylinder \cite{wait1955scattering} or spheroid\cite{asano1975light}, ellipsoid, elliptical infinite cylinder \cite{ren1997scattering}, etc. Furthermore, with exception of the sphere and the infinite circular cylinder, the calculable size of these theories is often limited to several tens of wavelengths\cite{xu2007theoretical}. This limitation is primarily due to the problem of numerical calculation of the involved special functions.

The numerical methods solve the scattering problems by discretizing the equations, for instance, the finite-difference time-domain \cite{taflove2005computational,2013FiniteDTD}, the discrete dipole approximation \cite{yurkin2011discrete,2019LirenxianScattering,chaumet2022discrete,zhou2023quantifying}, the T-Matrix \cite{2011wangjjNumerical,martin2019t,sun2019invariant,zhong2020t,2022MarstonScattering,huShuai2023}, etc.  They can deal with particles of any shape, but they are very costly in computation resources and time. The calculable size of particles exceeds difficultly few hundred wavelengths.

The approximate models are often employed to address the scattering of particles with irregular shapes, but their precision is often insufficient. Rayleigh model \cite{rayleigh1897v,vcivzmar2006optical} serves as a good approximation for particles much smaller than the wavelength. Conversely, the geometrical optics approximation (GOA) can only handle the scattering of large particles of dimension much larger than the wavelength. van de Hulst \cite{van1957light} has considered in details the interaction of light rays with a spherical particle. This analysis includes different properties of the light, such as the intensity, the optical path phase, the phase of focal lines. The extension of the GOA to the shaped beam scattering by both spherical and non-spherical particle has also been explored \cite{xu2006extension,yu2008geometrical,yu2009geometrical,li2009computation,muinonen1996light,grin2002scattering}. Nevertheless, the precision of the GOA is  limited and it is challenging to account for the divergence/convergence of a wave on the surface of the particle.

In conclusion, none of the previously mentioned theories, models or methods can handle the light scattering of large non-spherical particles with good precision. To overcome this obstacle, the Vectorial Complex Ray Model (VCRM) \cite{2011vectorial,ren2018vectorial,jiang2013theoretical} has been developed. In this model, the wavefront curvature (WFC) is introduced as an intrinsic property of light rays, which enables an easy description of wave divergence and convergence on the curved surface of the particle and the calculation of the phase due to focal lines through the wavefront equation. Since the WFC variation depends only on the local curvature of the particle surface and the calculation is done step by step, the VCRM can be applied to the scattering of light by large particles of any shape with smooth surface. The VCRM has been validated experimentally \cite{onofri2015experimental} and numerically \cite{yang2015comparison} in the cases of scattering in a symmetric plane of scatterer. The divergence factor is investigated in detail for the scattering of an infinite cylinder and an ellipsoid by combing the VCRM and the physical optics \cite{Zhang2023DivergenceFI}. It has been shown that the combination of the VCRM and the physical optics (PO) permits to explain clearly and to remedy the problem encountered in the classical Airy theory\cite{2022Airy}.

However, the three-dimensional light scattering of a non-spherical particle by the VCRM, a two-dimensional interpolation is unavoidable for the calculation of the total scattered field. Because of the irregularity of the emergent ray directions (azimuth and elevation angles) this task is very difficult. Though a line-by-line interpolation has been developed and applied to the scattering of a plane wave by a spheroidal particle \cite{DUAN2019106677}, a real liquid jet \cite{DUAN2017156,DUAN2019106677}, an oblate drop \cite{Duan2021Generalized} and the scattering of a Gaussian beam by a spherical particle\cite{Duan2023Numerical}, an alternative and flexible method is still necessary to deal with the scattering of particle of more complicated shape. This can also serve as a check to the results obtained by the interpolation. To this end, the Statistic Vectorial Complex Ray Model (SVCRM) is proposed by Rozé\cite{2019YangNumerical}. It has been shown that without consideration of the interference, the scattering patterns simulated with SVCRM agree well with the skeletons of the images obtained experimentally\cite{yang2018simulation}.

In this article, we will extend the SVCRM to its complete version, i.e. the different kinds of the phase shift in the interaction of a ray with the particle are counted and the complex amplitude of the scattered field is calculated. It permits, therefore, to predict the fine structure of scattering diagram including the interference effect.  We choose the pendent drop as example because it is a real large non-spherical particle encountered in many applications and it is easy to obtain experimentally a stable drop. The surface profile of the scatterer will be extracted from its image and described by an empiric polynomial function.

The rest of the paper is organized as follows. In Section \ref{sec:theoryModel}, the theoretical model and the related algorithm will be described. The simulated scattering patterns of pendent drops of different shape will be compared to the experimental ones in Section \ref{sec:results}. The scattering mechanism, especially the scattering patterns around rainbow angles, will be analyzed, with emphasis on the contribution of different orders of rays to different scattering regions. Section \ref{sec:conclusion} is devoted to the conclusions.

\section{Theoretical models}\label{sec:theoryModel}
The SVCRM is based on the VCRM. We will begin by a brief presentation of the VCRM, followed by a detailed description of the SVCRM and the algorithm.
\subsection{Vectorial Complex Ray Model}\label{sec:VCRM}
In VCRM\cite{2011vectorial,renJiang2012scattering,ren2012scattering,deschamps1972ray}, all waves are described by bundles of vectorial complex rays and each ray is characterized by its propagation direction -- wavevector $\bm{k}$, polarization state $X (\perp$ or $\|$ to the incident plane),  phase $\Phi$, amplitude $A_{X}$ as well as the wavefront curvature, which permits to deduce the amplitude and the phase due to the focal line of each individual ray at any point directly and simply.

\begin{figure}[ht!]
\centering
\includegraphics[width=6cm]{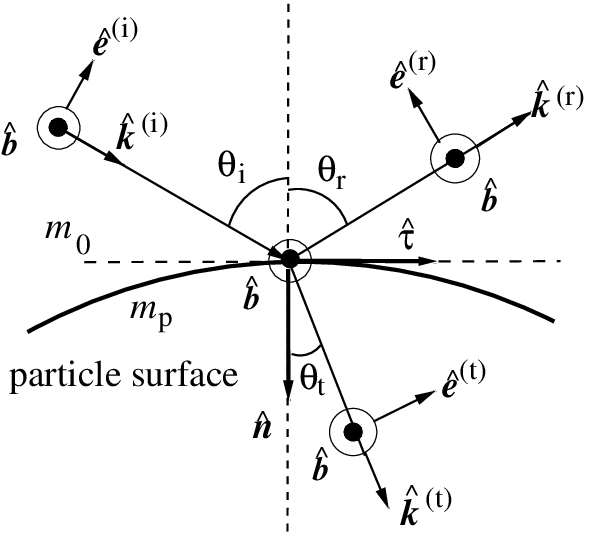} \\
\caption{Definition of the four bases used in SVCRM.}
\label{fig:BasesSVCRM}
\end{figure}

The directions of the reflected ray and the refracted ray are determined by the fact that the tangent components of their wavevector on the particle surface are continuous (Snell-Descartes law in vector form), i.e. the tangent components of the reflected ray $k_\tau^{(r)}$ and the refracted ray $k_\tau^{(t)}$ are equal to that of the incident ray $k_\tau^{(i)}$ \cite{2011vectorial,jiang2013theoretical}:
\begin{equation}\label{eq:k_tau-VCRM}
k_\tau^{(r)}=k_\tau^{(t)}=k_\tau^{(i)}
\end{equation}
where $k_\tau^{(r)}=k_\tau^{(i)}=\bm{k}^{(i)}\cdot\bm{\hat{\tau}}=k_0\sin \theta_{i}$, $k_\tau^{(t)}=\bm{k}^{(t)}\cdot\bm{\hat{\tau}}=mk_0\sin \theta_{t}$, $\bm{\hat{\tau}}$ being the unit vector in the incident plane and tangent to the particle surface, $\theta_{i}$ the incident angle, $\theta_{t}$ the refraction angle and $m=\frac{m_{p}}{m_{0}}$ the relative refractive index of the particle (Fig. \ref{fig:BasesSVCRM}), where $m_p$ is the refractive index of the particle and $m_0$ the refractive index of the surrounding..

In the VCRM, the Fresnel formulas are given as functions of the normal components of the wavevectors of the incident and the refracted waves, noted respectively by $k_n^{(i)}$ and $k_n^{(t)}$, as follows\cite{yurish2018advances}:
\begin{align} \label{eq:FCsV}
r_{\parallel} &= \dfrac{ m^{2}k_{n}^{(i)}-k_{n}^{(r)}}{m^{2} k_{n}^{(i)}+k_{n}^{(t)}}   &
t_{\parallel} &=  \dfrac{2 m k_{n}^{(i)}}{m^{2} k_{n}^{(i)}+k_{n}^{(t)}}  \nonumber  \\
r_{\perp} &= \dfrac{ k_{n}^{(i)}-k_{n}^{(r)}}{ k_{n}^{(i)}+k_{n}^{(t)}}  &
t_{\perp} &= \dfrac{2 k_{n}^{(i)}}{k_{n}^{(i)}+k_{n}^{(t)}}
\end{align}
with $k_{n}^{(i)}=\bm{k}^{(i)}\cdot\bm{\hat{n}}=k_0\cos \theta_{i}$, $k_{n}^{(t)}=\bm{k}^{(t)}\cdot\bm{\hat{n}}=mk_0\cos \theta_{t}$, $\bm{\hat{n}}$ being the normal of the particle surface. $k_n^{(i)}$ and $k_n^{(t)}$  can also be calculated without knowing the angles according to 
\begin{equation}
k_n^{(r)} = - k_n^{(i)}, \qquad
k_n^{(t)} = \sqrt{{k^{(t)}}^2-{k_\tau^{(i)}}^2}
\label{eq:k_n-VCRM}
\end{equation}

In the differential geometrical point of view, the curvature of a curved surface can be described by a $2 \times 2$ curvature matrix in a given base on the tangent surface. Consider a wave of wavefront curvature at the incident point described by the curvature matrix $\mathbf{Q}$ in the bases ($\bm{t}_1,\bm{t}_2$) arriving on a dioptric surface of curvature matrix $\mathbf{C}$ in the bases ($\bm{s}_1,\bm{s}_2$). The curvature matrix of the refracted or reflected wavefront $\mathbf{Q}'$ in its bases $(\bm{t}_1', \bm{t}_2')$ is given in the VCRM by the wavefront equation  \cite{2011vectorial}:
\begin{equation}\label{eq:WF_VCRM}
\left( \bm{k}' - \bm{k}\right)\cdot\bm{n}\mathbf{C} = k' \bm{\Theta}^{'T}\mathbf{Q}'\bm{\Theta}'-k\bm{\Theta}^{T}\mathbf{Q}\bm{\Theta}
\end{equation}
where $k$ and $k'$ are the wave numbers of the rays before and after interaction. The superscript $^T$ denotes the transposition of the matrix. In addition, $\bm{\Theta}$ and $\bm{\Theta}'$ are the projection matrices of the bases $(\bm{t}_1,\bm{t}_2 )$ and $(\bm{t}_1', \bm{t}_2')$ on the bases ($\bm{s}_1,\bm{s}_2$) respectively and given as:
\begin{equation}\label{eq:ThetaBA}
	\bm{\Theta} = \left[ \begin{array}{clcr}
		\bm{s}_1 \cdot \bm{t}_1 &   \bm{s}_1 \cdot \bm{t}_2 \\
		\bm{s}_2 \cdot \bm{t}_1 &   \bm{s}_2 \cdot \bm{t}_2
	\end{array}  \right],  \qquad
	\bm{\Theta'} =\left[ \begin{array}{clcr}
		\bm{s}_1 \cdot \bm{t}_1' &   \bm{s}_1 \cdot \bm{t}_2' \\
		\bm{s}_2 \cdot \bm{t}_1' &   \bm{s}_2 \cdot \bm{t}_2'
	\end{array}  \right]
\end{equation}

The divergence and the convergence of a wave can be calculated directly by the Gaussian curvatures of the wavefront surfaces step by step. So, the VCRM can be applied to the scattering of light by large particles of any shape with smooth surface.  This will be described in detail in the following section on the Statistical Vectorial Complex Ray Model (SVCRM).

\subsection{Statistical Vectorial Complex Ray Model}\label{sec:SVCRM}			
In statistical methods, the light wave is often presented by photons. However, in the SVCRM, the total field at given direction will be calculated by summation of all the ``photons'' arriving in the same tiny box and these photons have the same properties of rays in VCRM \cite{yang2018simulation}. Furthermore, the problem under study is the scattering of a continuous wave by a stable particle, so there is no temporal effect. So we will continue to use ``rays'' in the SVCRM as in the VCRM instead of ``photons''. In this section, the algorithm for the calculation of the physical properties of rays will be described in details.
								
\textbf{Four coordinate bases}: In the scattering of light by a particle of arbitrary shape, the direction of the incident wave, the polarization state of a wave, the principal directions of the particle surface and those of the wavefront change at each interaction. To describe these properties we define four normalized orthogonal bases at a intersection point as shown in Fig. \ref{fig:BasesSVCRM}.

The normal of the particle surface at a given interaction point $\bm{\hat{n}}$ is chosen such that the angle between $\bm{\hat{n}}$ and the incident wavevector $\bm{k}$ is acute (less than $\pi/2$), i.e. $\ubm{n}$ is inward at the first incident point and outward at all the other interaction points. The incidence plane is then defined by the propagation direction of the incident wave $\bm{k}^{(i)}$ and the normal of the surface $ \ubm{n}$. The unit vector normal to the incidence plane is defined by $\ubm{b} = \ubm{n}\times\ubm{\tau}$. 
To describe the properties of a ray, we define four normalized orthogonal bases in the incidence plane: $(\ubm{n},\ubm{\tau}, \ubm{b})$ for the particle surface, $(\bm{\hat{e}}^ {(i)}, \ubm{b},  \bm{\hat{k}}^ {(i)} )$ for the incident ray,  $( \ubm{e}^ {(r)},  \ubm{b}, \ubm{k}^{(r)} )$ for the reflected ray and $(\ubm{e}^{(t)}, \ubm{b}, \ubm{k}^{(t)})$ for the refracted ray. 

\textbf{Directions of emergent rays}:
The tangent and normal components of the reflected and refracted wavevectors are determined according, respectively, to  Eq. (\ref{eq:k_tau-VCRM}) and  Eq. (\ref{eq:k_n-VCRM}). The wavevector are given as :
								
\begin{equation}\label{eq:krt}
\bm{k}^{(r)} = k_\tau^{(i)}\ubm{\tau}+ k_n^{(r)}\ubm{n}, \qquad
\bm{k}^{(t)} = k_\tau^{(i)}\ubm{\tau}+ k_n^{(t)}\ubm{n}
\end{equation}
								
\textbf{Curvature matrix}: 
At an interaction point of a ray with the particle surface, the curvature matrix of the dioptric surface $\mathbf{C}$, the wavefront curvature matrices of the incident wave $\mathbf{Q}^{(i)}$, the reflected wave $\mathbf{Q}^{(r)}$ and the refracted wave $\mathbf{Q}^{(t)}$ can be given in their bases defined in Fig. \ref{fig:BasesSVCRM}, respectively, as:
\begin{equation}\label{eq:CQQ_GSVCRM}
\begin{split}
\mathbf{C} = \left[ \begin{array}{clcr}
\kappa_{ii}^{(s)} & \kappa_{ij}^{(s)} \\
\kappa_{ij}^{(s)}  & \kappa_{jj}^{(s)}
\end{array}  \right],  \quad
\bm{Q}^{(i)} =\left[ \begin{array}{clcr}
\kappa_{ii}^{(i)}  & \kappa_{ij}^{(i)} \\
\kappa_{ij}^{(i)}  & \kappa_{jj}^{(i)}
\end{array}  \right],  \\[2mm]
\bm{Q}^{(r)} =\left[ \begin{array}{clcr}
\kappa_{ii}^{(r)} & \kappa_{ij}^{(r)} \\
\kappa_{ij}^{(r)} & \kappa_{jj}^{(r)}
\end{array}  \right],\quad
\bm{Q}^{(t)} =\left[ \begin{array}{clcr}
\kappa_{ii}^{(t)} & \kappa_{ij}^{(t)} \\
\kappa_{ij}^{(t)} & \kappa_{jj}^{(t)}
\end{array}  \right]
\end{split}
\end{equation}
								
The projection matrix in Eq. (\ref{eq:ThetaBA}) are simple in this case:
\begin{equation}\label{eq:ThetaBA_SVCRM}
\bm{\Theta}^{(i)} = \left[ \begin{array}{clcr}
\cos \theta_{i} &   0 \\
	0 &   1
\end{array}  \right], \qquad
\bm{\Theta^{(r)}} =\left[ \begin{array}{clcr}
\cos \theta_{r} &   0 \\
	0 &   1
\end{array}  \right],\qquad
\bm{\Theta^{(t)}} =\left[ \begin{array}{clcr}
\cos \theta_{t} &   0 \\
0 &   1
\end{array}  \right]
\end{equation}

The wavefront curvatures of the reflected ray and the refracted ray are deduced from Eq. (\ref{eq:WF_VCRM}).
For the reflected ray, $k^{(r)}=k^{(i)}$ and $\bm{k}^{(r)}\cdot\ubm{n}=-\bm{k}^{(i)}\cdot\ubm{n}$.  The wavefront equation is given by:
\begin{equation}\label{eq:WFr_SVCRM}
k^{(r)} \bm{\Theta^{(r)T}}\mathbf{Q}^{(r)}\bm{\Theta^{(r)}}=k^{(i)}\bm{\Theta^{(i)T}}\mathbf{Q}^{(i)}\bm{\Theta^{(i)}}- 2 \bm{k}^{(i)}\cdot\ubm{n}\mathbf{C}
\end{equation}
And the wavefront equation for the refracted ray is:
\begin{equation}\label{eq:WFt_SVCRM}
k^{(t)}\bm{\Theta^{(t)T}} \mathbf{Q}^{(t)} \bm{\Theta^{(t)}} = k^{(i)} \bm{\Theta^{(i)T}} \mathbf{Q}^{(i)} \bm{\Theta^{(i)}} +(\bm{k}^{(t)}\cdot\ubm{n}-\bm{k}^{(i)}\cdot\ubm{n})\mathbf{C}
\end{equation}
								
However, Eqs. (\ref{eq:WFr_SVCRM}) and (\ref{eq:WFt_SVCRM}) give only the curvature matrices of the wavefronts just after the interaction point $M$. To continue the trace of the ray, we need the WFC at the next incident interaction point $M'$ . This can be done by using the relation of the wavefront curvature radii $R$ and $R'$ between two points $M$ and $M$  of distance $l$
\begin{equation}
	R' = R-l
\label{RpR}
\end{equation}
Here we use the convention that a positive curvature corresponds to a convergent wave.
We need therefore the two curvature radii $R_1$ and $R_2$ of the wavefront at $M$. As we know that a curvature matrix is diagonal when it is given in its principal direction bases and the two elements in the diagonal are the two curvatures and their inverses are the two curvature radii. The diagonalization of a $2\times2$ matrix is an easy task. Suppose that the two diagonalized curvature matrices  $\mathbf{Q}^{(r)}$ and $\mathbf{Q}^{(t)}$ are respectively:
\begin{equation}\label{eq:CQQ_DSVCRM}
\mathbf{Q}^{(r)} = \left[ \begin{array}{clcr}
\kappa_{1}^{(r)} & 0\\
	0 & \kappa_{2}^{(r)}
\end{array}  \right],   \qquad
\mathbf{Q}^{(t)} =\left[ \begin{array}{clcr}
\kappa_{1}^{(t)} & 0 \\
0 & \kappa_{2}^{(t)}
\end{array}  \right]
\end{equation}

The curvature radius is then given by $R=1/\kappa$ where $\kappa$ takes any value of the four elements in Eq. (\ref{eq:CQQ_DSVCRM}) and $R$ is the corresponding curvature radius. The curvature radius of the wave at a distance $l$ is given by Eq. (\ref{RpR}).

It is important to note that usually neither of the two principal directions of the wave incident on the a point is in the incident plane. Before the application of Eqs. (\ref{eq:CQQ_GSVCRM}) to (\ref{eq:WFt_SVCRM}) the curvature matrix of the incident wave must be transformed to the incident plane bases as described above. The curvature matrix of the particle surface should also be given in the base $(\ubm{n},\ubm{\tau},\ubm{b} )$.
								
\begin{figure}[ht!]
\centering
\includegraphics[width=6.5cm]{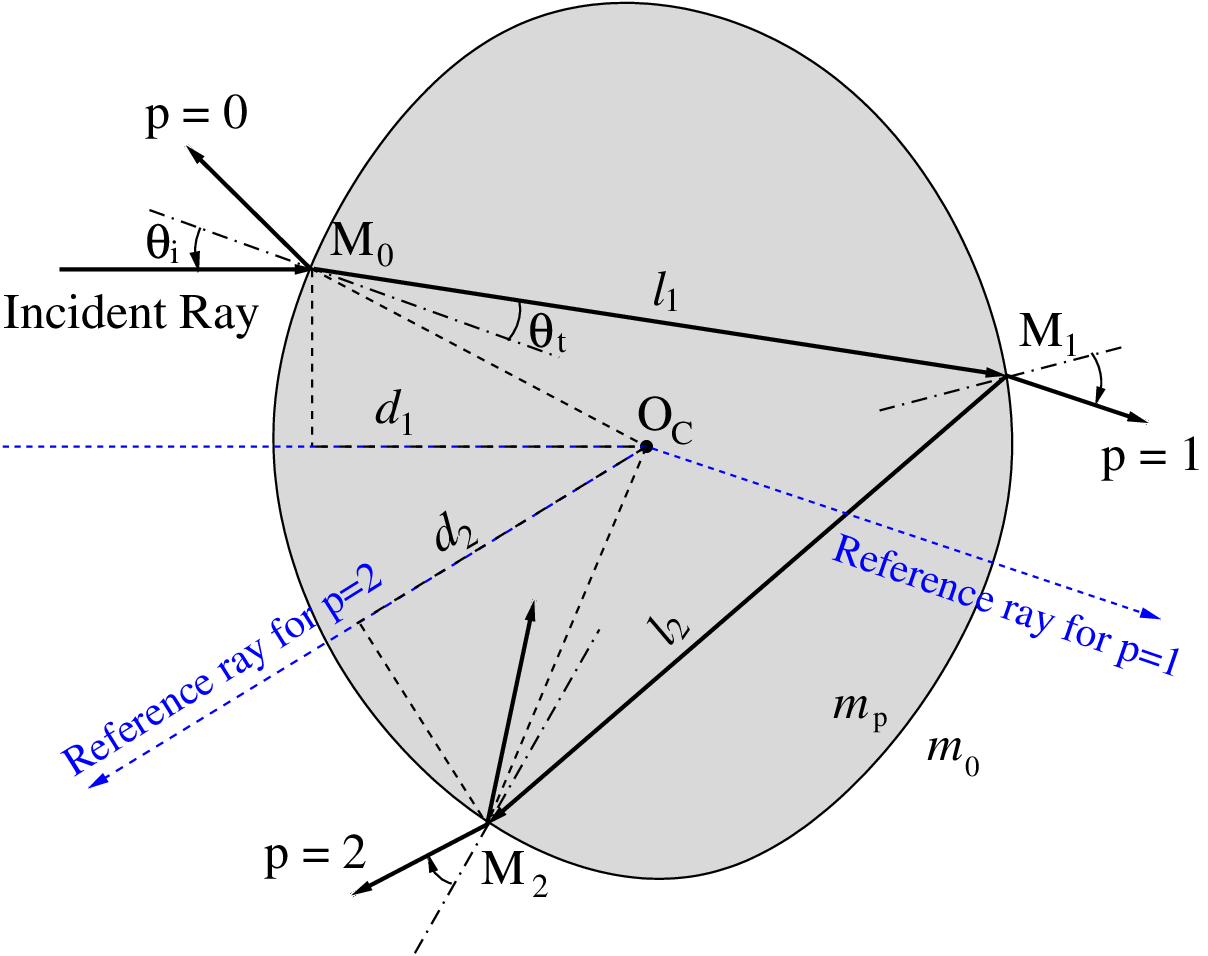}
\caption{Ray tracing in a non-spherical particle}\label{fig:DropletOP}
\label{fig:RaysInDrop}
\end{figure}

\textbf{Phase of a ray}: 							
The phases of the rays play a critical role in the study of interference phenomenon. In the SVCRM, we take as reference the ray which arrives at the center of the particle or the origin of the coordinate system in the same direction as the incident wave and goes out in the same direction as the emergent wave as if there is no particle (dashed blue lines in Fig. \ref{fig:RaysInDrop}). Then three kinds of phase shift relative to this reference are considered.
		
The first is the phase shift due to reflection. This is calculated directly according to the Fresnel coefficients \cite{2019YangNumerical,2020duan3Dphd}. For a transparent particle, if no total reflection occurs, only a phase shift $\pi$ is to be added to the reflected ray for the perpendicular polarization at any incident angle and for the parallel polarization if the incident angle is larger than the Brewster angle. There is no phase shift for the refracted ray whatever the polarization state. When the total reflection occurs, the phase shifts of the reflection and refraction rays depend on the incident angle and can be calculated directly:					
\begin{equation}\label{eq:PhiFCs}
{\Phi_{F,X}} = \arg(\epsilon_{X})
\end{equation}
where $\epsilon_X$ is the Fresnel reflection coefficient $r_X$ or $t_X$ given in Eq. (\ref{eq:FCsV}).

The second is the phase shift due to optical path which is computed in respect to the reference mentioned above. The optical path of the emergent ray of order $p=2$, for example, is calculated by $\Delta L_2=m(l_1+l_2)-(d_1+d_2)$ with $d_{1} = \overrightarrow{M_{0} O}_{C}  \cdot \bm{\hat{k}}^{(i)}$ and $d_{2} = \overrightarrow {O_{C} M}_{2} \cdot \bm{\hat{k}}^{(e)}$, $\bm{\hat{k}}^{(e)}$ being the normalized wavevector of the emergent wave.

For any order, let $l_i$ be the distance between two successive points of $i^{th}$ and $(i+1)^{th}$ interactions of a ray with the particle surface, the optical path of the ray of order $p$ can be calculated by
\begin{equation}\label{eq:alpha2}
\Delta L_{p} =  m \sum_{i=1}^{p} l_{i}-(d_1+d_p)
\end{equation}
where $d_{p} = \overrightarrow {O_{C} M}_{p} \cdot \bm{\hat{k}}^{(e)}$.
The phase shift induced by optical path is therefore 
\begin{equation}\label{eq:PhiOP2}
\Phi_{P,p} = \dfrac{2\pi}{\lambda} \Delta L_{p}
\end{equation}
where $\lambda$ is the wavelength in the surrounding medium.
				
The last one is the phase shift due to focal lines. At each passage of a focal line the phase advances by ${\pi}/{2} $. This corresponds to a sign change of a wavefront curvature. Therefore, the calculation of the phase due to the focal line in the SVCRM is just a matter to count the number of sign changes $N_f$ of the wavefront curvatures (a focal point is a crossing of two focal lines at the same point). The phase shift due to the focal lines can be calculated directly:
\begin{equation}\label{eq:PhaseFocalLine}
\Phi_f =  \frac{\pi}{2} \cdot N_{f}
\end{equation}													
It should be noted that the only purpose of introducing the WFC in the SVCRM is to compute accurately this phase shift.
						
In summary, the total phase shift of a emergent ray of order $p$ is given by
\begin{equation}\label{Eq:SVCRM-phaseshifts}
\Phi_{X,p} =  \Phi_{F,X,p} +\Phi_{P,p} + \Phi_{f,p}
\end{equation}
The index $p$ is added to indicate that the phase shifts are calculated for the emergent ray of that order.		

\textbf{Amplitude of a ray}:  In the SVCRM, the light source, plane wave in our study, is simulated by a homogenously distributed rays and the total scattered intensity in a given direction is calculated by the summation of the contribution of all the rays arriving in a small box $\Delta\theta \cdot  \Delta\phi$ in that direction. To count correctly the interference effect, the two polarizations must be treated separately. However, for a non-spherical particle, the direction of the incident plane changes at each interaction point and the cross polarization occurs. So we cannot give a general formula but it is not difficult in practice to calculate complex amplitude by using the relation between the complex amplitude before the $(i+1)^{th}$ and $i^{th}$ interactions given as 
\begin{equation}\label{eq:Ap_i}
\tilde{A}_{X,i+1}=\tilde{A}_{X,i}|\epsilon_{X,i}|e^{i\Phi_{X,i,i+1}}
\end{equation}							
where $X$ stands for the polarization state on the incident plane at the $i^{th}$ interaction point. $\Phi_{X,i,i+1}$ is the phase shift between the two points.

Since we are interested only in the scattering intensity in far field, the complex electric field of each emergent ray can be decomposed in the directions parallel and perpendicular to the scattering plane -- plane defined by the direction of the incident wave and the observation point. Let $\bm{\hat{e}}_\perp$ and $\bm{\hat{e}}_\parallel$ be the unit vectors perpendicular and parallel to the scattering plane, the two components of the amplitude of a ray can then be expressed in this base by

\begin{equation}\label{eq:Ap_paraperp}
\begin{split}
{\tilde{A}_{\parallel,p}}  = \tilde{A}_{i,p} \bm{\hat{e}_{i,p}}  \cdot  \bm{\hat{e}_{\parallel}}  + \tilde{A}_{j,p} \bm{\hat{e}_{j,p}} \cdot \bm{\hat{e}_{\parallel}}  \\
{\tilde{A}_{\perp,p}} = \tilde{A}_{i,p} \bm{\hat{e}_{i,p}} \cdot \bm{\hat{e}_{\perp}}  +  \tilde{A}_{j,p} \bm{\hat{e}_{j,p}}  \cdot \bm{\hat{e}_{\perp}}
\end{split}
\end{equation}							
								
Let the collection box be defined by two angle steps  $\Delta\phi$ and $\Delta\theta$ at given  direction $(\phi, \theta)$.
The scattered intensity is the total energy flux collected by the box divided by the solid angle of the box $\Delta\Omega= \Delta\phi \cdot \Delta\theta \cdot \cos \theta$ ($\theta$ here is the complementary angle of the polar angle in the standard spherical coordinate system).

\begin{equation}\label{eq:I_SVCRM}
I(\theta,\phi) = \dfrac{\tilde{A}_{\parallel} \cdot {\tilde{A}_{\parallel}}^{*} + \tilde{A}_{\perp} \cdot {\tilde{A}_{\perp}}^{*}}{\Delta\Omega}
\end{equation}
where $\tilde{A}_{\parallel}=\sum \tilde{A}_{\parallel,p}$ and $\tilde{A}_{\perp}=\sum\tilde{A}_{\perp,p}$. The summation here means all the rays (rays of different ordres $p$ and from the same order) arriving in the box.

\section{Simulation, validation and discussion}\label{sec:results}
The method described in the previous sections will be applied to the simulation of the scattering patterns of a pendent drop. The results are compared to the experimental measurement to validate the algorithm. We will be interested particularly in the intensity distribution around the rainbow angles because it is the most sensible to the deformation of the particle.
						
\subsection{Experimental setup and acquired data} \label{subsec:dropletexp}
								
\begin{figure}[ht!]
\centering
\includegraphics[width=7cm]{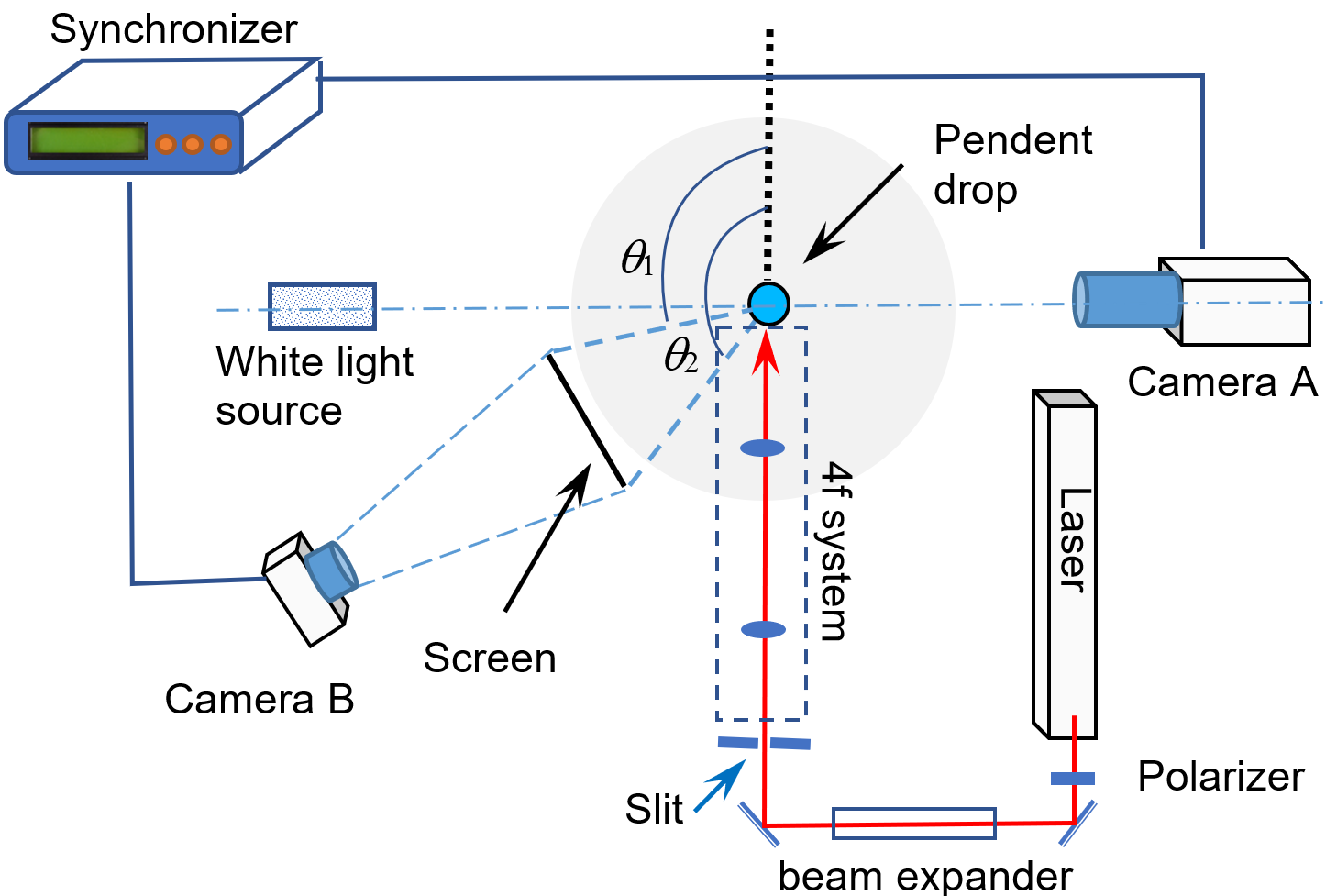}
\caption{Experimental setup}\label{fig:Experimental_setup}
\end{figure}

The experimental setup in Fig. \ref{fig:Experimental_setup} is built up in order to obtain the drop image and the scattering pattern simultaneously. The pendent water drop is generated by a circular tube of internal diameter 0.5 mm connected to a tank of 10 liters with a fine flowmeter in order to obtain a drop of wanted size. A high-quality laser (He-Ne JDSU Laser, Model 1145P, class 3B) emits a beam of diameter 0.7 mm at wavelength $\lambda=0.6328~\mu$m. After passing through a polarizer, the beam is expanded to about 1.5 mm and then illuminates the pendent drop. The image of the drop is recorded by Camera A, a CCD camera JAI 10-bit model TM-4200CL of resolution $2048 \times 2048$ pixels. The scattering pattern is registered by Camera B, A HAMAMATSU 14-bit model C9100-02 CDD camera with a resolution of $1000 \times 1000$ pixels. A white screen is used to obtain the scattering pattern. The two cameras are connected to a BNC pulse generator synchronizer to record simultaneously images by both cameras. Moreover, a slit and 4f system are used to obtain a beam slice to illuminate a part of the drop. This permits to isolate the contribution of the light from different parts of the drop and to investigate the mechanism of the scattering in details. 

\begin{figure}[ht!]
\centering
\includegraphics[width=5.5cm]{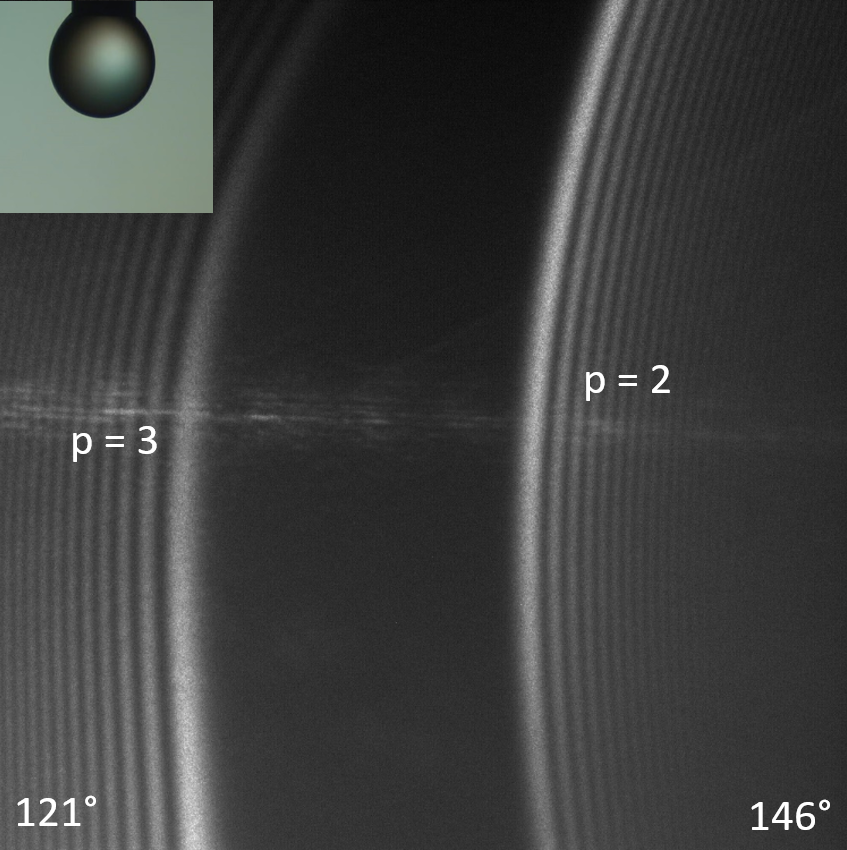}\hspace{5mm}
\includegraphics[width=5.5cm]{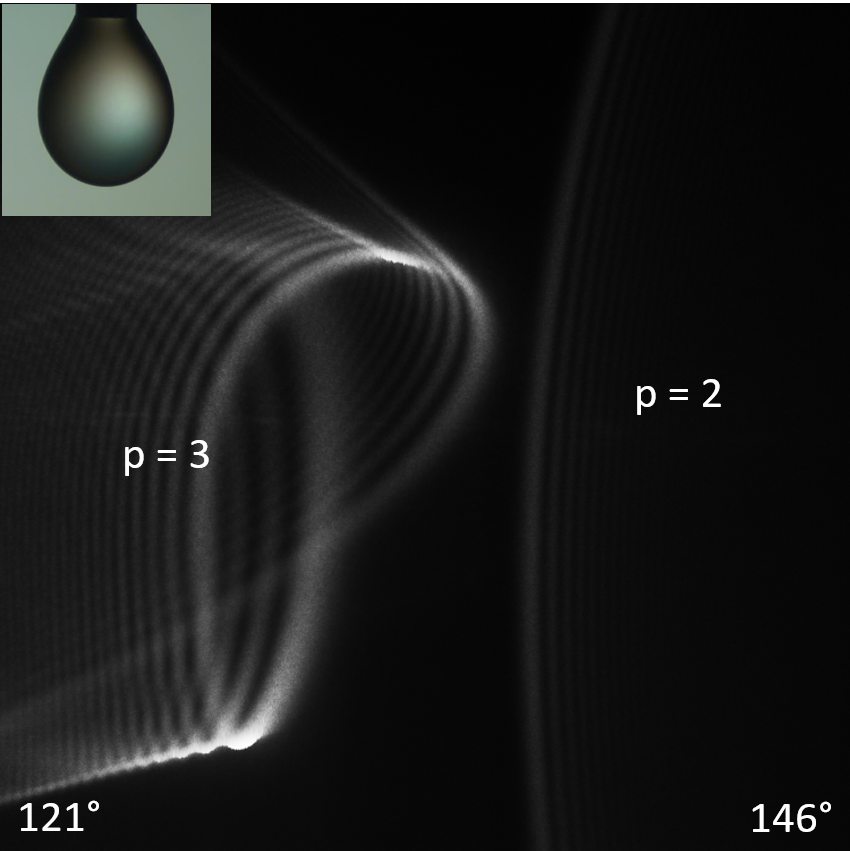}\\
(a) \hspace{5.5cm} (b) 
\caption{Experimental images of two pendent drops (in the top-left coner taken by the camera A) and their scattering patterns around the rainbow angles with a perpendicularly polarized plane wave.}
\label{fig:Exp_droplets}
\end{figure}

The drop images and the corresponding scattering patterns near the first ($p=2$) and the second ($p=3$) rainbows are shown in Fig. \ref{fig:Exp_droplets} for two pendent drops of typical shapes. We find that when the drop is small (Fig. \ref{fig:Exp_droplets}(a)) the shape is almost spherical and the scattering pattern is similar to that of a sphere, i.e. the fringes in the first and the second rainbows are almost parallel. When the drop is large, it is elongated (Fig. \ref{fig:Exp_droplets}(b)), the form of the first rainbow remains almost the same but the second rainbow is completely twisted and its intensity is dominant.

\begin{figure}[ht!]
\centering
\includegraphics[width=5.5cm]{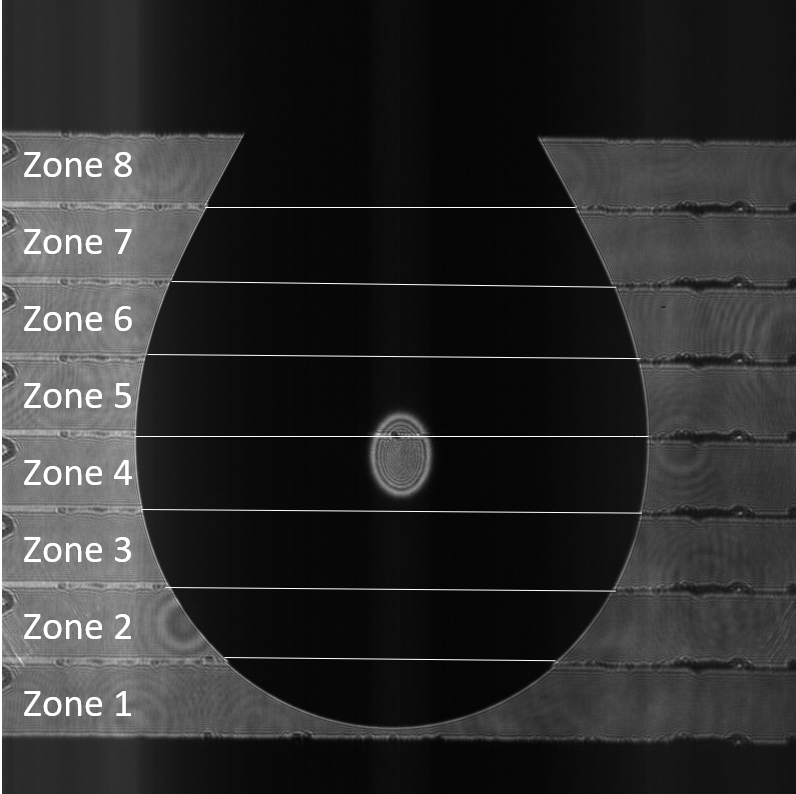}\hspace{5mm}
\includegraphics[width=5.5cm]{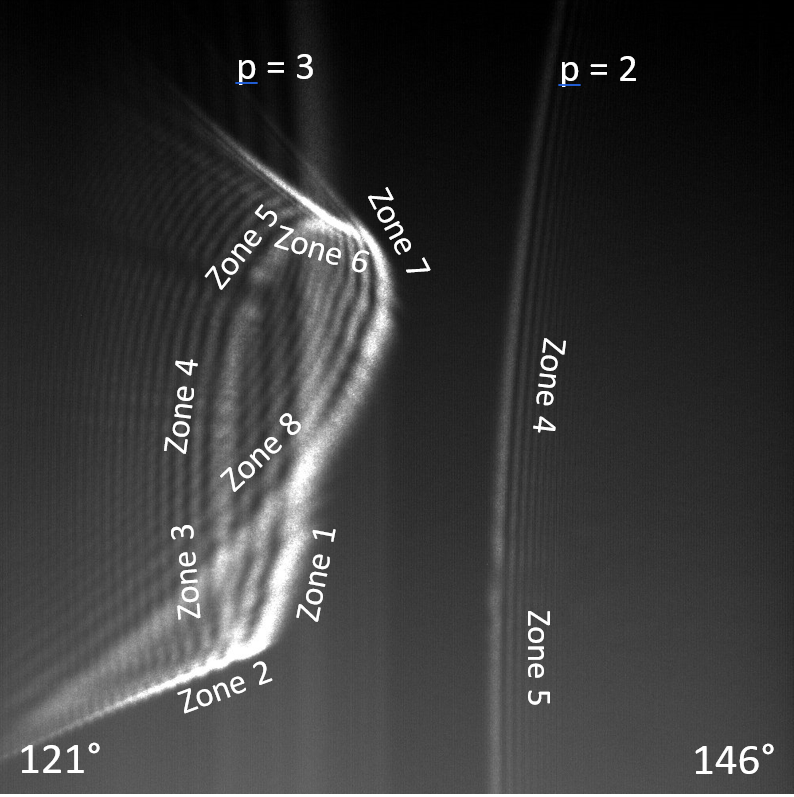} \\
 (a) \hspace{5.5cm}         (b)    \\
\caption{Illumilated zones (left) and corresponding scattering patterns (right). The images shown here are composed of 8 recorded images for each individual zone.}
\label{fig:8zones}
\end{figure}

To understand the scattering mechanism, the drop is illuminated by a beam slice. We show in Fig. \ref{fig:8zones} the illumination zones (a) and the corresponding scattering patterns (b). The same light source has been used for the particle image and the scattering patterns, i.e. the camera A is in front of the 4f system (see Fig. \ref{fig:Experimental_setup}). By moving the beam slice from the bottom to the top of the drop we obtain eight scattering diagrams.  Fig. \ref{fig:8zones} (b) is the scattering diagram composed of the eight recorded scattering patterns. We see that in the first  rainbow, the scattering patterns near the equatorial plane (zones 4 and 5) are contributions of the light near the equatorial plane (zone 5 and 4, so inversed). But the scattering pattern of the second rainbow near the equatorial plane contains the light from all the drop (zones 1 to 8). These will be investigated by the numerical simulation of the SVCRM. 
 
\subsection{Drop profile description} 
\begin{figure}[ht!]
\centering
\includegraphics[width=5.5cm] {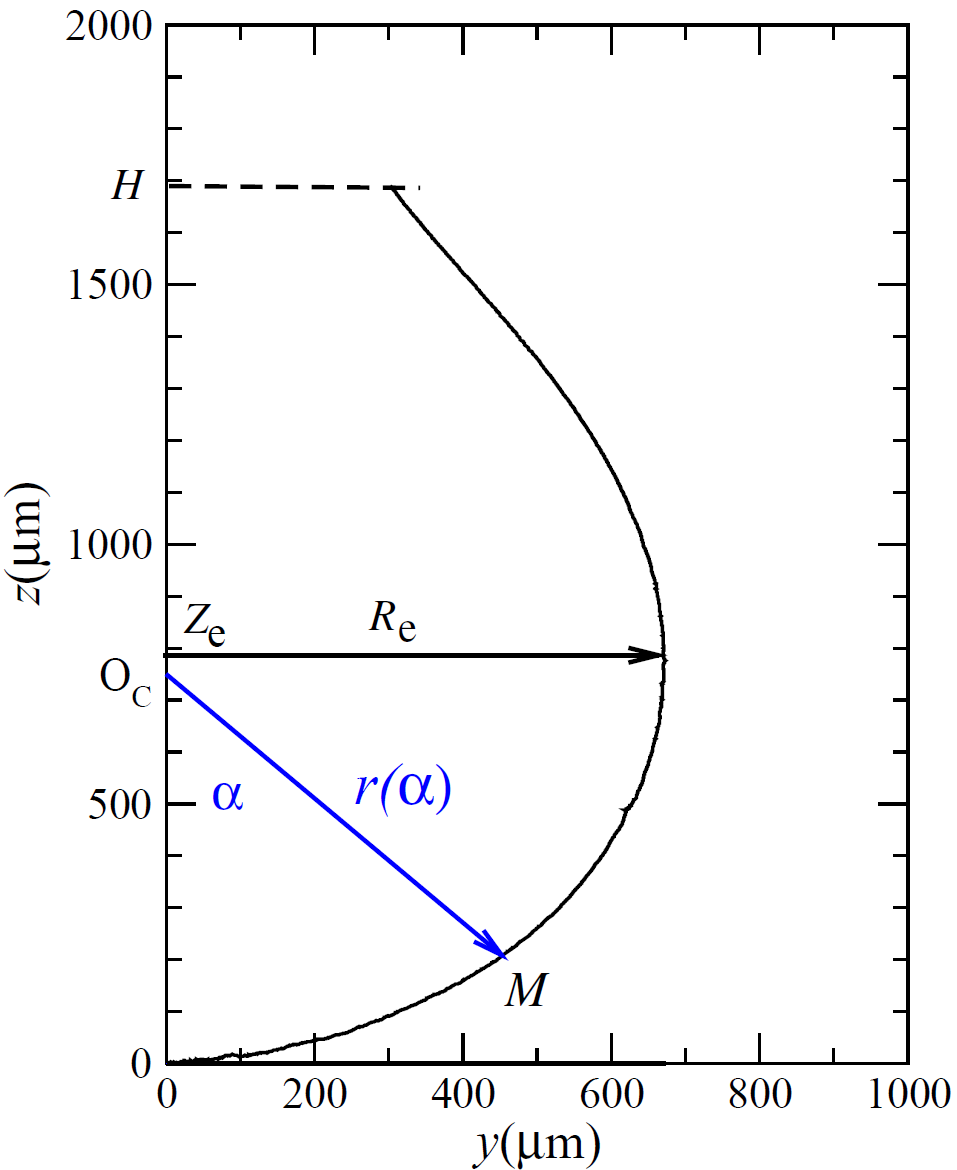}
\caption{Drop profile for data fitting.}
\label{fig:DropProfile}
\end{figure}

In order to simulate the scattering of the particle, we need a mathematical description of its profile.  The pendant drop being circularly symmetric, its shape can be described by the profile in a symmetric plane with a one-variable function. We choose to express the distance of a point on the surface as function of the angle with the vertical direction $\alpha$ i.e. $r(\alpha)$ (Fig. \ref{fig:DropProfile}). In our calculation, a $10^{th}$ degree polynomial is adopted.

\begin{equation}\label{eq:r_theta}
r(\alpha)=\sum_{k=0}^{10} {a_{k}\alpha^{k}}
\end{equation}
The coefficients $a_{k}$ are determined by the least-square fitting of the data extracted from the contour of the particle image. These coefficients of the two typical drops in Fig. \ref{fig:Exp_droplets} are given in Tab. \ref{tab:Coefak}. In this data fitting, the bottom pixel in the image is chosen as the origin of $z$ axis and $O_C$ is located at 750 $\mu$m. Because of the symmetry of the drop, the tangent plane at the bottom of the drop is horizontal, the derivative of the function $r(\alpha$) is zero, and so coefficient $a_1$ is required to be zero in the least-square fitting algorithm. In addition, the size of a drop is characterized by its height $H$, its radius $R_{e}$ in the equatorial plane and the position of its equatorial plane $Z_{e}$ along $z$ axis. These parameters are also given in Tab. \ref{tab:Coefak}.

\begin{table}[ht!]
\centering
\caption{Coefficients $a_k$ of the polynomial function of Eq. (\ref{eq:r_theta}) for the two drops in Fig. \ref{fig:Exp_droplets} and their parameters.}
{\small
\begin{tabular}{|c|r|r||c|r|r|}
\hline
drop  & a  & b	  & drop & a	&   b\\
\hline\hline
$ a_{0} $ 	& $ 749.62142$ & $ 748.94172 $ &  $ a_{7} $ 	& $ 2902.87230$	& $-1474.38707 $		 \\	
\hline	
$a_1$ 		& 0 					& 0  						&$ a_{8} $ 	& $-971.01492 $  & $ 499.55719$ \\
\hline 
$ a_{2} $ 	& $-429.66991$ & $85.95885 $	&  $ a_{9} $  	& $ 180.01080$	& $-92.76912 $	 \\	
\hline
$ a_{3} $ 	& $ 1524.47565$ & $ -702.79373 $	&	$ a_{10}$ & $-14.19570$ & $ 7.26156 $	 \\	
\hline
$ a_{4} $ 	& $-3923.68613 $	& $1864.65631 $  & 	$ R_{e} $($\mu$m)   & $547.35$	    & $706.73 $	 \\	
\hline
$ a_{5} $ 	& $ 5848.04927$	& $ -2818.77887 $	&  $ H $($\mu$m)	    & $1151.18$ 	  & $1787.41$ \\	
\hline
$ a_{6} $ 	& $-5245.52714$	& $2601.86539 $	& $Z_{e}$($\mu$m)    & $589.20$ 	 	& $762.94$  \\
\hline
\end{tabular}
}
\label{tab:Coefak}
\end{table}	
								
\begin{figure}[ht!]
\centering
\includegraphics[width=5.5cm]{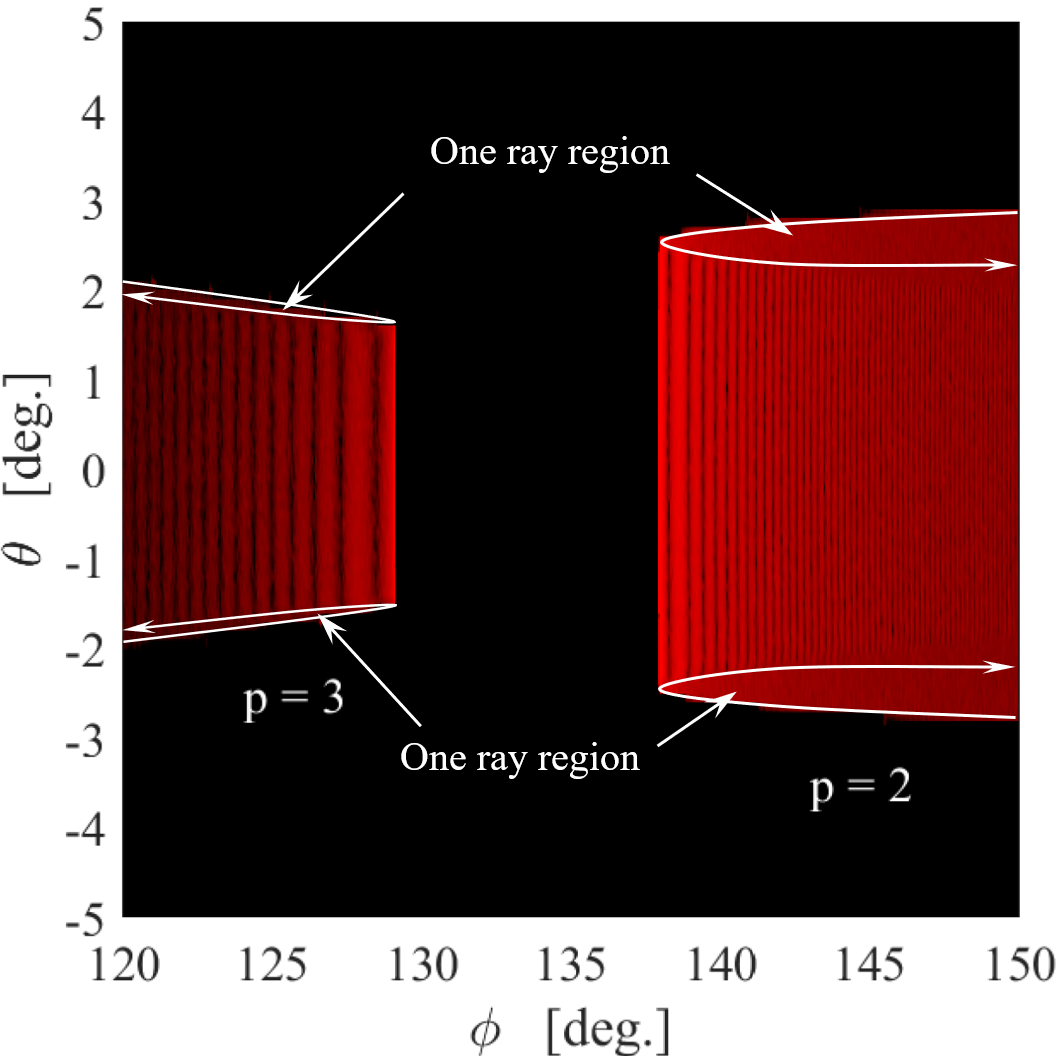} \hspace{5mm}
\includegraphics[width=5.5cm]{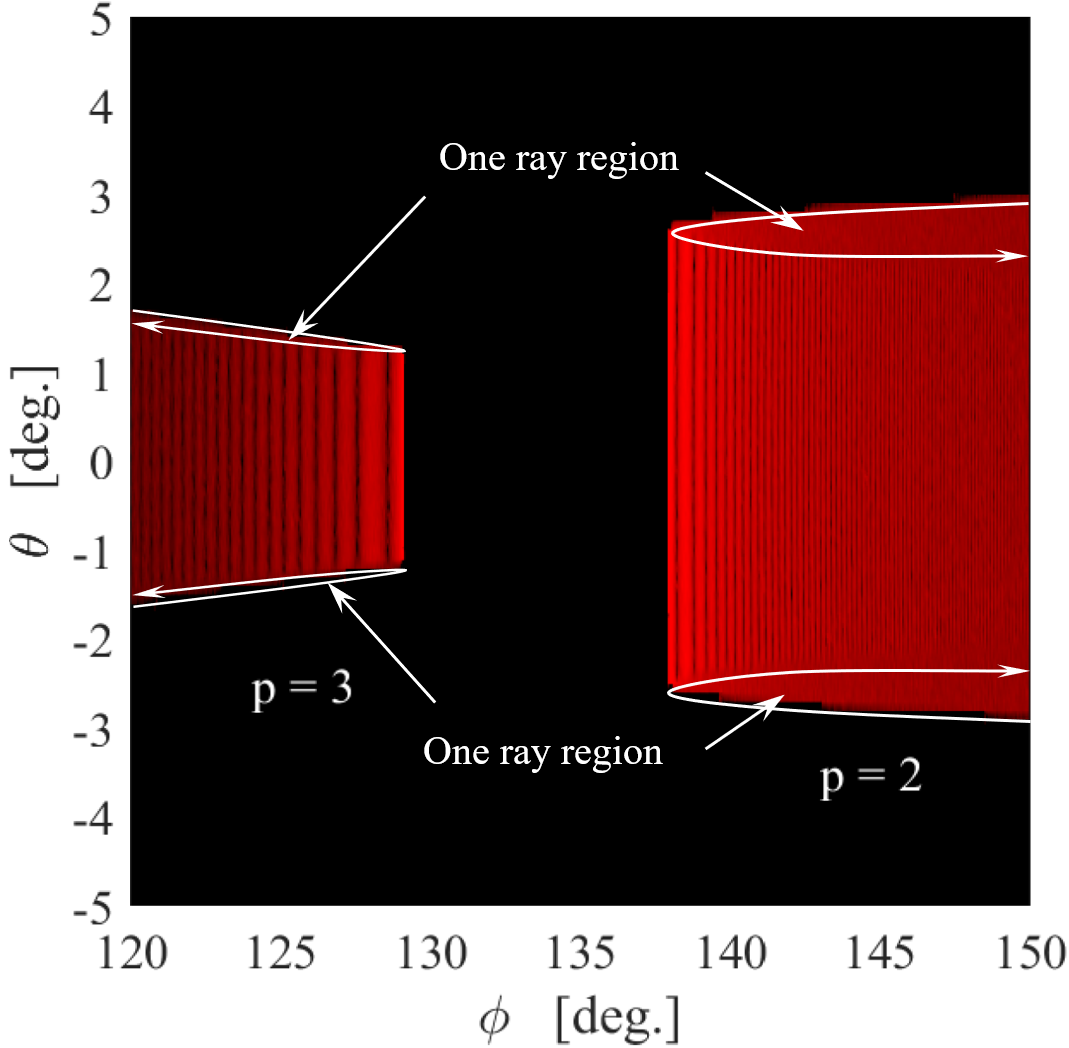} \\
\hspace{5mm} (a) \hspace{5.5cm}         (b)    \\
\caption{Scattering patterns near the equatorial plane of the pendent drops with the parameters in Tab. \ref{tab:Coefak}. The height and the width of illumination region are respectively $Z_{s} \in \left[-0.05R_{e}, 0.05R_{e} \right]$ and $y_{s}\in \left[-R_{e}, R_{e} \right]$. The total number of incident rays is $N=4 \times 10^{8}$ and the detection steps $\left(\Delta\phi,\Delta\theta\right)=\left(0.02^\circ, 0.1^\circ\right)$.}
\label{fig:equat-ab}
\end{figure}

\subsection{Simulation and discussion}
The incident laser beam is much larger than the diameter of the pendent drop. So in our simulation, we consider a particle illuminated by the incident plane wave propagating along $x$ axis (Fig. \ref{fig:DropProfile}). The coordinates $y,z$ of the rays are generated with a standard random number generator (in Fortran) homogeneously distributed in the illumination region. The observation direction is characterized by the azimutal angle $\phi$ in the $xy$ (horizontal) plane and the elevation angle $\theta$ relative to $xy$ plane. The relative refractive index of the water is taken to be 1.333.

In the SVCRM calculation, the size of detection box $(\Delta\phi, \Delta\theta)$ is important for accounting the interference effect. It must be sufficiently small so that the phases of the emergent rays arriving in the same detection box be nearly constant and sufficiently big in order to collect enough rays for a given total number of rays. To estimate the detection steps $\Delta\phi$, we start by simplifying the pendent drop to an equivalent spheroid of semi-axes $a=b=R_{e}$ and $c$ equal to the vertical curvature of the drop at $z=Z_e$. The scattering pattern in the equatorial plane of such spheroid can be calculated by using the VCRMEll2D software\cite{VCRMEll2D}. This step enables us to estimate rapidly the detection angle step $\Delta\phi$, which is $0.02^\circ$ for the two drops under study. 

The simulation is firstly done for the drops (Tab. \ref{tab:Coefak}) illuminated by a beam slice on the equatorial plane. The results are shown in Fig. \ref{fig:equat-ab}. In this case, the variation of the intensity in the vertical direction is weak, so the angular step $\Delta\theta$ is taken to be 5 times larger than $\Delta\phi$. We note that the drop (b) is more elongated than the drop (a) so the scattering pattern is less stretched. We find also a zone ($\phi>137.9^\circ$) for $p=2$ where there is no interference because only one ray of $p=2$ arrives in this region for a given values of $(\phi,\theta$).   

\begin{figure}[ht!]
\centering
\includegraphics[width=5.5cm] {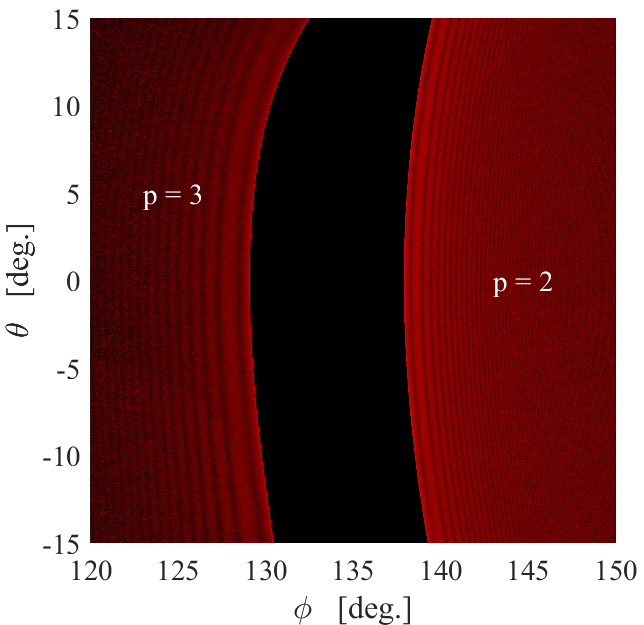} \hspace{5mm}
\includegraphics[width=5.5cm] {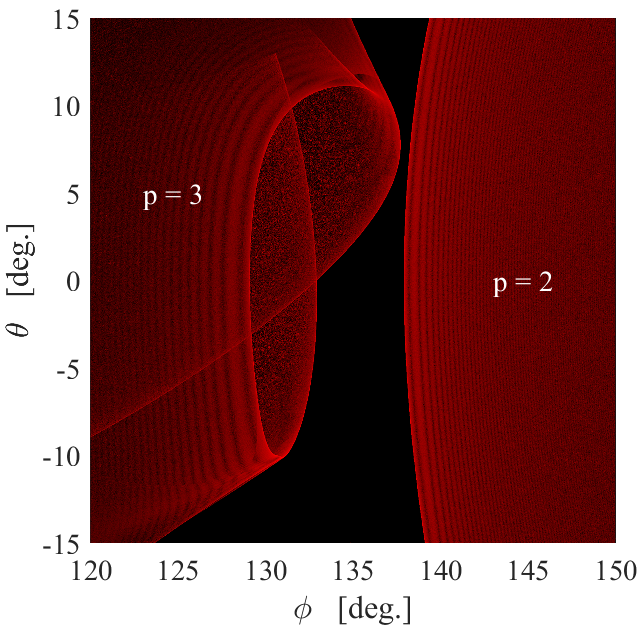} \\
\hspace{5mm}(a)  \hspace{5.5cm} (b) 
\caption{Scattering patterns around the rainbow angles of the two drops in Fig. \ref{fig:Exp_droplets}. The illumination region is $Z_{s}\in \left[0, H \right]$and $y_{s}\in \left[-R_{e}, R_{e} \right]$. The total number of incident rays is $N=2 \times 10^{9}$ and the detection steps $\left(\Delta\phi, \Delta\theta\right)$=$\left(0.02^\circ, 0.02^\circ\right)$.}
\label{fig:ScatPatterns}
\end{figure}

Fig. \ref{fig:ScatPatterns} shows the scattering patterns near the horizontal plane around the first and the second rainbow angles of the two drops when they are illuminated fully by a plane wave. We note firstly that the scattering patterns are in good agreement with those obtained by experiment in Fig. \ref{fig:Exp_droplets}. In this simulation, the angular step $\Delta\theta$ is the same as $\Delta\phi$ because the variation of the intensity in the vertical direction is also rapid as in the horizontal direction. The computation time for a scattering pattern is about 7 hours on a workstation of 16 processors and 3 Go RAM for each process.

\begin{figure}[h!]
\centering
\includegraphics[width=5.6cm]{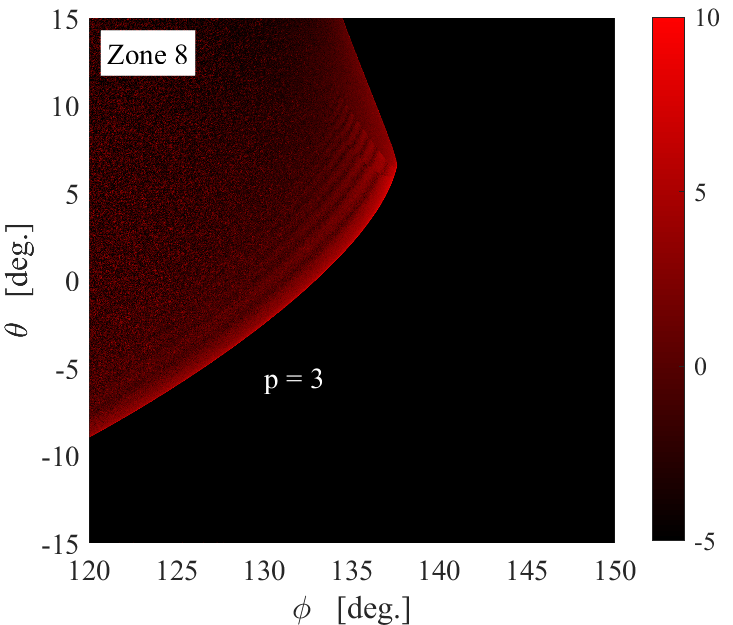} 
\includegraphics[width=5.6cm]{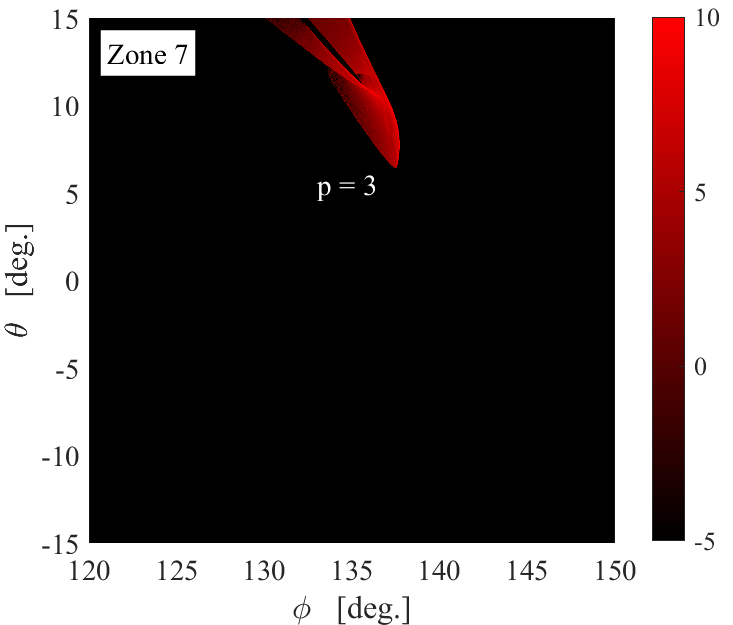}\\
\includegraphics[width=5.6cm]{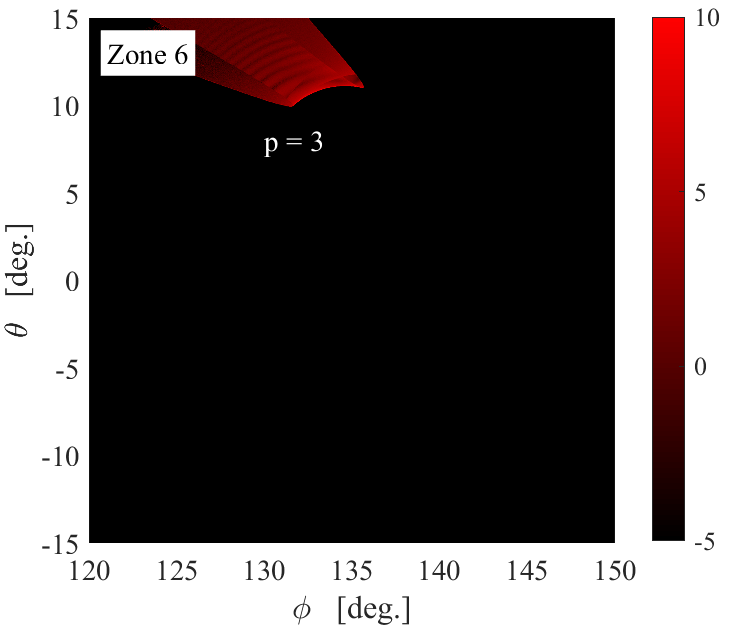} 
\includegraphics[width=5.6cm]{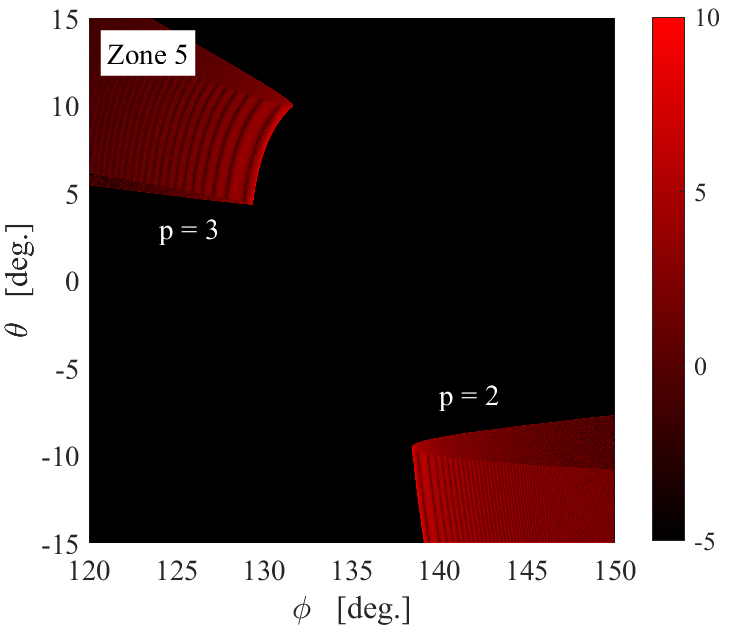}  \\
\includegraphics[width=5.6cm]{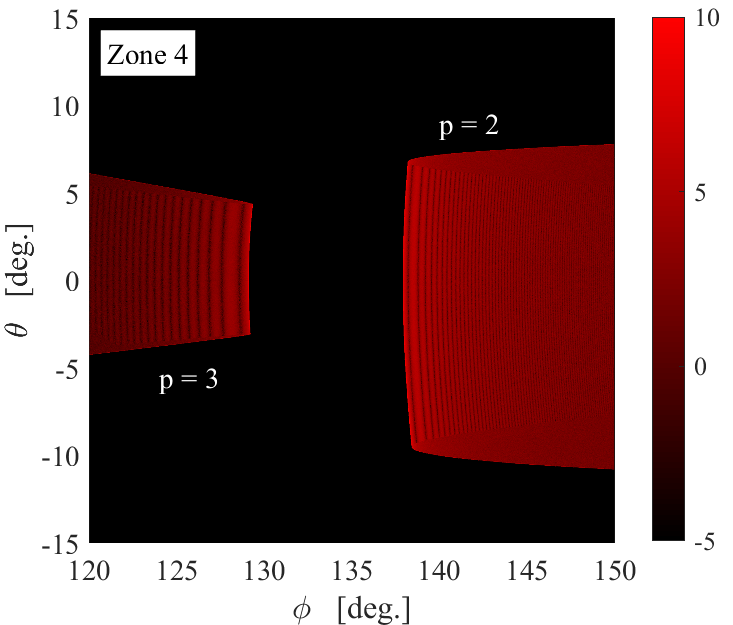} 
\includegraphics[width=5.6cm]{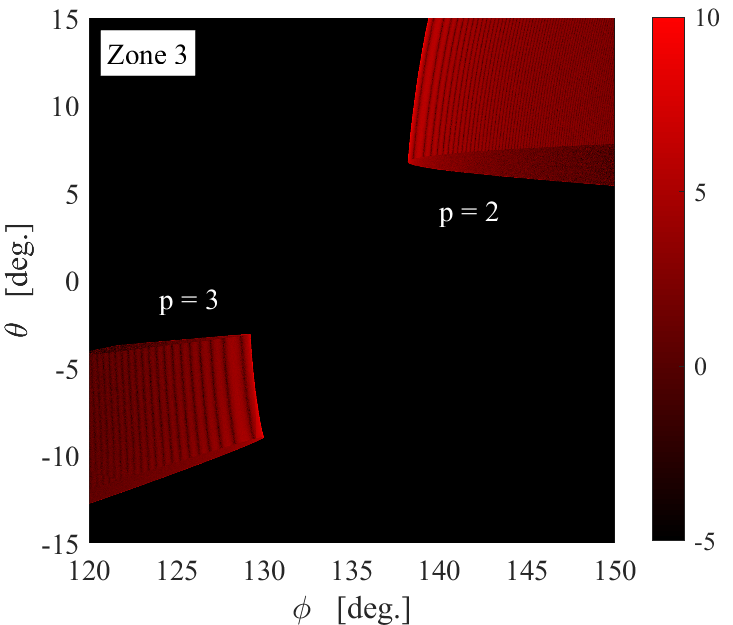}\\
\includegraphics[width=5.6cm]{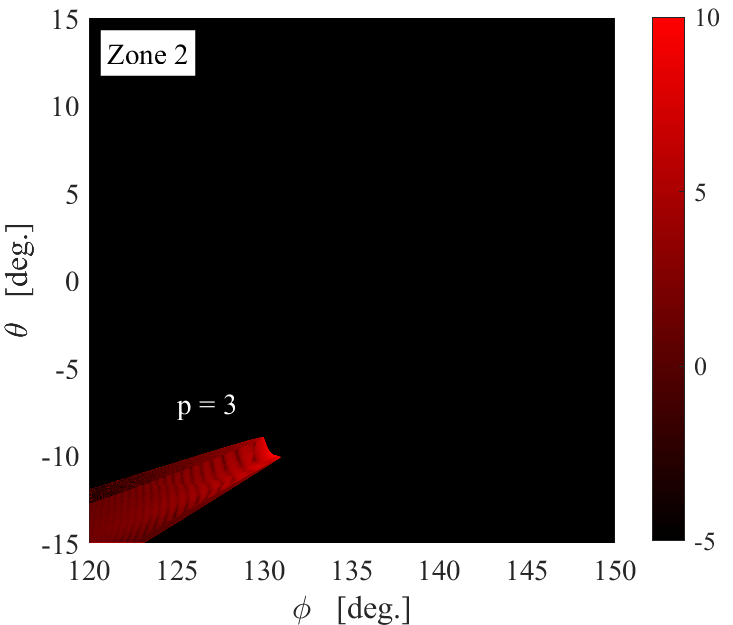} 
\includegraphics[width=5.6cm]{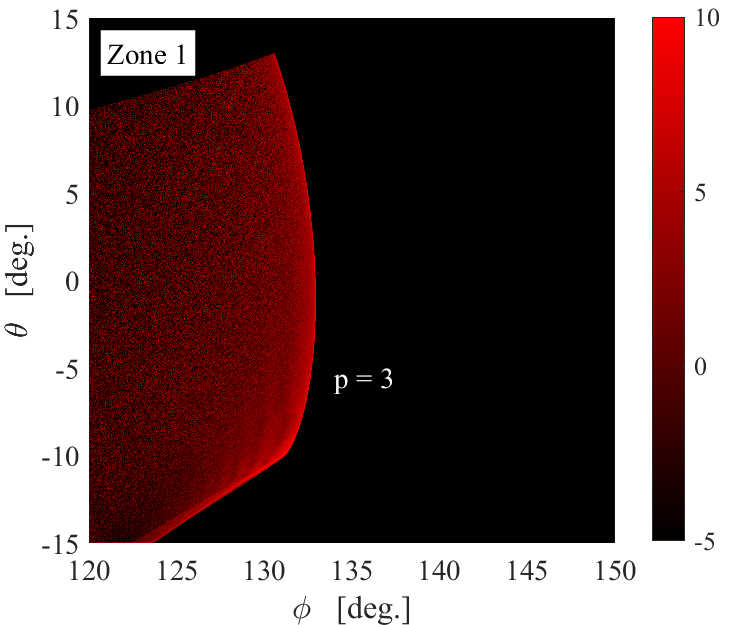}
\caption{Scattering patterns the pendent drop (b) illuminated by a beam slice at different position. The illumination region $i_z$ is defined by $z_{s}\in \left[(i_z-1)H/8), i_zH/8 \right]$ with $i_z=1-8$ and $y_{s}\in \left[-R_{e}, R_{e} \right]$. The total number of incident rays $N = 2 \times 10^{9}$ and the detection steps are $\left(\Delta\phi, \Delta\theta\right)$=$\left(0.02^\circ, 0.02^\circ\right)$. }
\label{fig:incregs_b1A8_s80}
\end{figure}

We find that the scattering pattern of the drop (b) forms a crossed loop with a bifurcation at the top of the loop (Figs. \ref{fig:Exp_droplets} (b) and \ref{fig:ScatPatterns} (b)). By illuminating the drop with beam slice, we have also isolated experimentally the contribution of the light from different zones. The wave being described by bundles of rays, the SVCRM is very suitable to simulate this effect and investigate the mechanism of the scattering. Fig. \ref{fig:incregs_b1A8_s80} depicts  the scattering patterns of the pendent drop (b) when it is illuminated by a beam slice of height equal to 1/8 of the drop height. To be clear, the scattering pattern of each zone is shown separately. In the simulation, only the rays of orders $p=2$ and $3$ are considered. As in Fig. \ref{fig:incregs_b1A8_s80}, we find the one ray-region more or less visible at the border of all zones for the two orders $p=2,3$ because at the rainbow angle the ray direction returns as shown in Fig. \ref{fig:equat-ab}. The bifurcation structure is the contribution of the light in the zone 7 from the neck of the drop.
																
\section{Conclusions}\label{sec:conclusion}
The Statistical Vectorial Complex Ray Model and the algorithm are presented and validated by the comparison of the simulated scattering patterns and those obtained experimentally for a pendent drop of different size and form. It is shown that by taking the interference phenomena into consideration, the SVCRM can be applied to the scattering of a large non-spherical particle. It permits also to investigate the scattering mechanism of the interaction of the light with a object of any form with smooth surface. However, because the scattering pattern of a pendent drop is very sensible to its size and shape, especially for an elongated drop, and it is very difficult to calibrate precisely the absolute angles, we have not been able to do the quantitative comparison between the experimental patterns and the simulated results. We are trying to solve the problem by two independent methods: the first consists of improving the experimental setup and the second is to construct a database with different parameters so that to determine the unknowns of the drop by machine learning.

Compared to the VCRM, the SVCRM is more flexible and can be applied easily to an object of complex form, of convex or concave surface for example. But it is more time compsuming. It is expected that the SVCRM will be applied in the development of the measurement techniques in the particle characterization and scattering properties of different objects. 


\section*{Funding}
Natural Science Basic Research Program of Shaanxi (Program No. 2023-JC-QN-0074); Fundamental Research Funds for the Central Universities (Program No. XJS222705). This work has been also partially supported by the CRIANN (Centre Régional Informatique et d'Applications Numériques de Normandie).


\section*{Disclosures}
The authors declare no conflicts of interest.

\section*{Data availability}
Data underlying the results presented in this paper are not publicly available at this time but may be obtained from the authors upon reasonable request.

\bibliography{svcrm}

\begin{thebibliography}{10}
\newcommand{\enquote}[1]{``#1''}

\bibitem{mayinger2013optical}
F.~Mayinger, \emph{Optical measurements: techniques and applications} (Springer
  Science \& Business Media, 2013).

\bibitem{sirohi2018optical}
R.~Sirohi, \emph{Optical methods of measurement: wholefield techniques} (CRC
  Press, 2018).

\bibitem{Zhao_2023}
H.~Zhao, X.~Wang, R.~Wang, D.~Hua, K.~Li, and F.~Ji, \enquote{Optimal static
  light scattering detection angle for particulate matter size and
  concentration measurement,} {\protect\JournalTitle{Measurement Science and
  Technology}} \textbf{34}, 125802 (2023).

\bibitem{mishchenko1999light}
M.~I. Mishchenko, J.~W. Hovenier, and L.~D. Travis, \emph{Light scattering by
  nonspherical particles: theory, measurements, and applications} (Elsevier,
  1999).

\bibitem{MISHCHENKO2009808}
M.~I. Mishchenko, \enquote{Electromagnetic scattering by nonspherical
  particles: A tutorial review,} {\protect\JournalTitle{Journal of Quantitative
  Spectroscopy and Radiative Transfer}} \textbf{110}, 808--832 (2009). Light
  Scattering: Mie and More Commemorating 100 years of Mie's 1908 publication.

\bibitem{2013Electromagneticnon}
W.~T. Rhodes, \emph{Electromagnetic Wave Scattering on Nonspherical Particles:
  Basic Methodology and Simulations} (Electromagnetic Wave Scattering on
  Nonspherical Particles: Basic Methodology and Simulations, 2013).

\bibitem{hulst1981light}
H.~van~de Hulst, \emph{Light scattering by small particles} (Dover
  Publications, InC. New York, 1981).

\bibitem{born1959principles}
M.~Born and E.~Wolf, \emph{Principles of optics: electromagnetic theory of
  propagation, interference, and diffraction of light} (Macmillan, 1959).

\bibitem{gouesbet1988light}
G.~Gouesbet, B.~Maheu, and G.~Gr{\'e}han, \enquote{Light scattering from a
  sphere arbitrarily located in a {Gaussian} beam, using a {Bromwich}
  formulation,} {\protect\JournalTitle{JOSA A}} \textbf{5}, 1427--1443 (1988).

\bibitem{Barton88}
J.~P. Barton, D.~R. Alexander, and S.~A. Schaub, \enquote{Internal and
  near-surface electromagnetic fields for a spherical particle irradiated by a
  focused laser beam,} {\protect\JournalTitle{J. Appl. Phys.}} \textbf{64},
  1632--1639 (1988).

\bibitem{book2017GLMT}
G.~Gouesbet and G.~Gréhan, \emph{Generalized Lorenz-Mie Theories} (2017),
  springer ed.

\bibitem{xu2007theoretical}
F.~Xu, K.~Ren, G.~Gouesbet, X.~Cai, and G.~Grehan, \enquote{Theoretical
  prediction of radiation pressure force exerted on a spheroid by an
  arbitrarily shaped beam,} {\protect\JournalTitle{Physical Review E}}
  \textbf{75}, 026613 (2007).

\bibitem{wait1955scattering}
J.~R. Wait, \enquote{Scattering of a plane wave from a circular dielectric
  cylinder at oblique incidence,} {\protect\JournalTitle{Canadian journal of
  physics}} \textbf{33}, 189--195 (1955).

\bibitem{asano1975light}
S.~Asano and G.~Yamamoto, \enquote{Light scattering by a spheroidal particle,}
  {\protect\JournalTitle{Applied optics}} \textbf{14}, 29--49 (1975).

\bibitem{ren1997scattering}
K.~Ren, G.~Gr{\'e}han, and G.~Gouesbet, \enquote{Scattering of a {Gaussian}
  beam by an infinite cylinder in the framework of generalized {Lorenz--Mie}
  theory: formulation and numerical results,} {\protect\JournalTitle{JOSA A}}
  \textbf{14}, 3014--3025 (1997).

\bibitem{taflove2005computational}
A.~Taflove and S.~C. Hagness, \emph{Computational electrodynamics: the
  finite-difference time-domain method} (Artech house, 2005).

\bibitem{2013FiniteDTD}
W.~Sun, G.~Videen, Q.~Fu, S.~Tanev, and J.~Huang, \enquote{Finite-difference
  time-domain solution of light scattering by arbitrarily shaped particles and
  surfaces,} {\protect\JournalTitle{Springer Berlin Heidelberg}}  (2013).

\bibitem{yurkin2011discrete}
M.~A. Yurkin and A.~G. Hoekstra, \enquote{The discrete-dipole-approximation
  code {ADDA}: capabilities and known limitations,}
  {\protect\JournalTitle{Journal of Quantitative Spectroscopy and Radiative
  Transfer}} \textbf{112}, 2234--2247 (2011).

\bibitem{2019LirenxianScattering}
R.~Zhang, Jiaming~Li, \enquote{Scattering of an airy light-sheet by a
  non-spherical particle using discrete dipole approximation,}
  {\protect\JournalTitle{Journal of Quantitative Spectroscopy and Radiative
  Transfer}} \textbf{225} (2019).

\bibitem{chaumet2022discrete}
P.~C. Chaumet, \enquote{The discrete dipole approximation: A review,}
  {\protect\JournalTitle{Mathematics}} \textbf{10}, 3049 (2022).

\bibitem{zhou2023quantifying}
C.~Zhou, X.~Han, and L.~Bi, \enquote{Quantifying the coherent backscatter
  enhancement of non-spherical particles with discrete dipole approximation,}
  {\protect\JournalTitle{Optics Express}} \textbf{31}, 24183--24193 (2023).

\bibitem{2011wangjjNumerical}
J.~Wang, G.~Gréhan, Y.~Han, S.~Saengkaew, and G.~Gouesbet, \enquote{Numerical
  study of global rainbow technique: sensitivity to non-sphericity of
  droplets,} {\protect\JournalTitle{Experiments in Fluids}} \textbf{51},
  149--159 (2011).

\bibitem{martin2019t}
T.~Martin, \enquote{T-matrix method for closely adjacent obstacles,}
  {\protect\JournalTitle{Journal of Quantitative Spectroscopy and Radiative
  Transfer}}  (2019).

\bibitem{sun2019invariant}
B.~Sun, L.~Bi, P.~Yang, M.~Kahnert, and G.~Kattawar, \emph{Invariant Imbedding
  T-matrix method for light scattering by nonspherical and inhomogeneous
  particles} (Elsevier, 2019).

\bibitem{zhong2020t}
H.~Zhong, L.~Xie, and J.~Zhou, \enquote{T-matrix formulation of electromagnetic
  wave scattering by charged non-spherical scatterers,}
  {\protect\JournalTitle{Journal of Quantitative Spectroscopy and Radiative
  Transfer}} \textbf{247}, 106952 (2020).

\bibitem{2022MarstonScattering}
P.~L. Marston and M.~I. Mishchenko, \enquote{Scattering by relatively small
  oblate spheroidal drops of water in the rainbow region: T-matrix results and
  geometric interpretation,} {\protect\JournalTitle{Journal of Quantitative
  Spectroscopy and Radiative Transfer}} \textbf{283}, 108142-- (2022).

\bibitem{huShuai2023}
H.~Shuai, L.~Shulei, Z.~Qingwei, and L.~Lei, \enquote{Dimension-variable
  invariant imbedding (dviim) t-matrix computational method for the light
  scattering simulation of atmospheric nonspherical particles,}
  {\protect\JournalTitle{Opt. Express}} \textbf{31}, 10052--10069 (2023).

\bibitem{rayleigh1897v}
L.~Rayleigh, \enquote{V. {On} the incidence of aerial and electric waves upon
  small obstacles in the form of ellipsoids or elliptic cylinders, and on the
  passage of electric waves through a circular aperture in a conducting
  screen,} {\protect\JournalTitle{The London, Edinburgh, and Dublin
  Philosophical Magazine and Journal of Science}} \textbf{44}, 28--52 (1897).

\bibitem{vcivzmar2006optical}
T.~{\v{C}}i{\v{z}}m{\'a}r, M.~{\v{S}}iler, and P.~Zem{\'a}nek, \enquote{An
  optical nanotrap array movable over a milimetre range,}
  {\protect\JournalTitle{Applied Physics B}} \textbf{84}, 197--203 (2006).

\bibitem{van1957light}
H.~Van~de Hulst, \enquote{Light {Scattering by Small Particle}, new york: John
  wiley \& sons,} {\protect\JournalTitle{Inc.}} pp. 114--130 (1957).

\bibitem{xu2006extension}
F.~Xu, K.~F. Ren, and X.~Cai, \enquote{Extension of geometrical-optics
  approximation to on-axis {Gaussian} beam scattering. {I}. by a spherical
  particle,} {\protect\JournalTitle{Applied optics}} \textbf{45}, 4990--4999
  (2006).

\bibitem{yu2008geometrical}
H.~Yu, J.~Shen, and Y.~Wei, \enquote{Geometrical optics approximation of light
  scattering by large air bubbles,} {\protect\JournalTitle{Particuology}}
  \textbf{6}, 340--346 (2008).

\bibitem{yu2009geometrical}
H.~Yu, J.~Shen, and Y.~Wei, \enquote{Geometrical optics approximation for light
  scattering by absorbing spherical particles,} {\protect\JournalTitle{Journal
  of Quantitative Spectroscopy and Radiative Transfer}} \textbf{110},
  1178--1189 (2009).

\bibitem{li2009computation}
X.~Li and X.~Han, \enquote{Computation of on-axis {Gaussian} beam scattering by
  nonuniform glass microbeads using a geometrical-optics approach,}
  {\protect\JournalTitle{Journal of Optics A: Pure and Applied Optics}}
  \textbf{11}, 105703 (2009).

\bibitem{muinonen1996light}
K.~Muinonen, T.~Nousiainen, P.~Fast, K.~Lumme, and J.~Peltoniemi,
  \enquote{Light scattering by {Gaussian} random particles: ray optics
  approximation,} {\protect\JournalTitle{Journal of Quantitative Spectroscopy
  and Radiative Transfer}} \textbf{55}, 577--601 (1996).

\bibitem{grin2002scattering}
E.~Grin’ko and Y.~G. Shkuratov, \enquote{The scattering matrix of transparent
  particles of random shape in the geometrical optics approximation,}
  {\protect\JournalTitle{Optics and Spectroscopy}} \textbf{93}, 885--893
  (2002).

\bibitem{2011vectorial}
K.~F. Ren, F.~Onofri, C.~Rozé, and T.~Girasole, \enquote{Vectorial complex ray
  model and application to two-dimensional scattering of plane wave by a
  spheroidal particle,} {\protect\JournalTitle{Optics letters}} \textbf{36},
  370--372 (2011).

\bibitem{ren2018vectorial}
K.~F. Ren and C.~Roz{\'e}, \emph{Vectorial Complex Ray Model for Light
  Scattering of Nonspherical Particles}, vol.~1 (IFSA Publishing, S. L, 2018).

\bibitem{jiang2013theoretical}
K.~Jiang, \enquote{Theoretical study of light scattering by an elliptical
  cylinder,} Ph.D. thesis, Texas A\&M University, USA, 2013. (2013).

\bibitem{onofri2015experimental}
F.~R. Onofri, K.~F. Ren, M.~Sentis, Q.~Gaubert, and C.~Pelc{\'e},
  \enquote{Experimental validation of the vectorial complex ray model on the
  inter-caustics scattering of oblate droplets,} {\protect\JournalTitle{Optics
  express}} \textbf{23}, 15768--15773 (2015).

\bibitem{yang2015comparison}
M.~Yang, Y.~Wu, X.~Sheng, and K.~F. Ren, \enquote{Comparison of scattering
  diagrams of large non-spherical particles calculated by {VCRM} and {MLFMA},}
  {\protect\JournalTitle{Journal of Quantitative Spectroscopy and Radiative
  Transfer}} \textbf{162}, 143--153 (2015).

\bibitem{Zhang2023DivergenceFI}
C.~Zhang, C.~Roz{\'e}, and K.~F. Ren, \enquote{Divergence factor in ray model
  and its application to light scattering by a large nonspherical dielectric
  particle,} {\protect\JournalTitle{IEEE Antennas and Wireless Propagation
  Letters}}  (2023).

\bibitem{2022Airy}
C.~Zhang, Claude.Rozé, and K.~F. Ren, \enquote{Airy theory revisited with the
  method combining vectorial complex ray model and physical optics,}
  {\protect\JournalTitle{Optics letters}} \textbf{47}, 2149--2152 (2022).

\bibitem{DUAN2019106677}
Q.~Duan, X.~Han, S.~Idlahcen, and K.~{Fang Ren}, \enquote{Three-dimensional
  light scattering by a real liquid jet: Vcrm simulation and experimental
  validation,} {\protect\JournalTitle{Journal of Quantitative Spectroscopy and
  Radiative Transfer}} \textbf{239}, 106677 (2019).

\bibitem{DUAN2017156}
Q.~Duan, R.~Zhong, X.~Han, and K.~F. Ren, \enquote{Influence of spatial
  curvature of a liquid jet on the rainbow positions: Ray tracing and
  experimental study,} {\protect\JournalTitle{Journal of Quantitative
  Spectroscopy and Radiative Transfer}} \textbf{195}, 156--163 (2017).
  Laser-light and Interactions with Particles 2016.

\bibitem{Duan2021Generalized}
Q.~Duan, F.~R.~A. Onofri, X.~Han, and K.~F. Ren, \enquote{Generalized rainbow
  patterns of oblate drops simulated by a ray model in three dimensions,}
  {\protect\JournalTitle{Opt. Lett.}} \textbf{46}, 4585--4588 (2021).

\bibitem{Duan2023Numerical}
Q.~Duan, F.~Onofri, X.~E. Han, and K.~F. Ren, \enquote{Numerical implementation
  of three-dimensional vectorial complex ray model and application to rainbow
  scattering of spheroidal drops.} {\protect\JournalTitle{Optics express}}
  \textbf{31 21}, 34980--35002 (2023).

\bibitem{2019YangNumerical}
R.~Yang, \enquote{Numerical simulation of light scattering of a pendent droplet
  by statistic vectorial complex ray model,} Ph.D. thesis, University of Rouen
  Normandy, 2013. (2019).

\bibitem{yang2018simulation}
R.~Yang, C.~Roz{\'e}, S.~Idlahcen, and K.~F. Ren, \enquote{Simulation of light
  scattering by a pendent drop with statistic vectorial complex ray model,}
  {\protect\JournalTitle{Sensors \& Transducers}} \textbf{226}, 71--76 (2018).

\bibitem{renJiang2012scattering}
K.~F. Ren, K.~Jiang \emph{et~al.}, \enquote{Scattering of an arbitrary shaped
  object by using vectorial complex ray model,} in \emph{ISAPE2012,}  (IEEE,
  2012), pp. 837--841.

\bibitem{ren2012scattering}
K.~Ren, C.~Roz{\'e}, and T.~Girasole, \enquote{Scattering and transversal
  divergence of an ellipsoidal particle by using vectorial complex ray model,}
  {\protect\JournalTitle{Journal of Quantitative Spectroscopy and Radiative
  Transfer}} \textbf{113}, 2419--2423 (2012).

\bibitem{deschamps1972ray}
G.~A. Deschamps, \enquote{Ray techniques in electromagnetics,}
  {\protect\JournalTitle{Proceedings of the IEEE}} \textbf{60}, 1022--1035
  (1972).

\bibitem{yurish2018advances}
S.~Y. Yurish, \emph{Advances in Optics Reviews 1} (Lulu. com, 2018).

\bibitem{2020duan3Dphd}
Q.~Duan, \enquote{On the three-dimensional light scattering by a large
  nonspherical particle based on vectorial complex ray model,} Ph.D. thesis,
  University of Xidian and University of Rouen Normandy, 2020. (2020).

\bibitem{VCRMEll2D}
K.~F. Ren, \enquote{{VCRMEll2D}, a free software to predict the scattering of a
  plane wave by an ellipsoid in the plane of symmetry,}
  {http:}//amocops.univ-rouen.fr/en/content/download.

\end{thebibliography}
\end{document}